\documentclass[journal]{IEEEtran}
\usepackage[ansinew]{inputenc}
\usepackage{amssymb}

\usepackage{tikz}
\usepackage{pgfplots}
\usetikzlibrary{shapes,decorations,arrows,decorations.markings,shapes.geometric,positioning,chains,patterns}
\usetikzlibrary{pgfplots.groupplots}

\pgfplotsset{compat=1.11}
\newcommand{\globalwidth}{0.8*\columnwidth}
\newcommand{\globalheight}{3cm}

\ifCLASSINFOpdf
\else
\fi

\usepackage{amsmath}
\usepackage{mathtools}
\usepackage{gensymb}
\usepackage{siunitx}
\usepackage{nicefrac}							
\usepackage{booktabs}
\usepackage{hhline}
\usepackage{multirow}
\usepackage{cite}

\usepackage{etoolbox}
\AtBeginEnvironment{eqnarray}{\setlength{\arraycolsep}{0.0em}}

\usepackage{bm}
\newcommand{\m}[1]{\bm{#1}} 

\makeatletter
\newcommand{\subalign}[1]{%
	\vcenter{%
		\Let@ \restore@math@cr \default@tag
		\baselineskip\fontdimen10 \scriptfont\tw@
		\advance\baselineskip\fontdimen12 \scriptfont\tw@
		\lineskip\thr@@\fontdimen8 \scriptfont\thr@@
		\lineskiplimit\lineskip
		\ialign{\hfil$\m@th\scriptstyle##$&$\m@th\scriptstyle{}##$\crcr
			#1\crcr
		}%
	}
}
\makeatother

\usepackage{amsthm}
\newtheorem{problem}{Problem}
\newtheorem{theorem}{Theorem}
\newtheorem{lemma}{Lemma}

\newtheorem{definition}{Definition}
\newtheorem{assumption}{Assumption}
\newtheorem{remark}{Remark}

\makeatletter
\renewcommand*\env@matrix[1][\arraystretch]{%
  \edef\arraystretch{#1}%
  \hskip -\arraycolsep
  \let\@ifnextchar\new@ifnextchar
  \array{*\c@MaxMatrixCols c}}
\makeatother

\begin{document}
%
\title{Inverse Dynamic Games Based on Maximum Entropy Inverse Reinforcement Learning}
%
%
%

\author{Jairo~Inga,
        Esther~Bischoff,
        Florian~K\"opf,
        and~Sören~Hohmann
\thanks{J. Inga, E. Bischoff, F. K\"opf and S. Hohmann are with the Institute of Control Systems, Karlsruhe Institute of Technology (KIT), Karlsruhe, Germany (e-mail: {jairo.inga, esther.bischoff, florian.koepf, soeren.hohmann}@kit.edu)}
}

\maketitle

\begin{abstract}
We consider the inverse problem of dynamic games, where cost function parameters are sought which explain observed behavior of interacting players. Maximum entropy inverse reinforcement learning is extended to the N-player case in order to solve inverse dynamic games with continuous-valued state and control spaces. We present methods for identification of cost function parameters from observed data which correspond to (i) a Pareto efficient solution, (ii) an open-loop Nash equilibrium or (iii) a feedback Nash equilibrium. Furthermore, we give results on the unbiasedness of the estimation of cost function parameters for each arising class of inverse dynamic game. The applicability of the methods is demonstrated with simulation examples of a nonlinear and a linear-quadratic dynamic game.
\end{abstract}


%
\IEEEpeerreviewmaketitle

\section{Introduction}
%
%
%
%
Dynamic game theory provides a useful mathematical tool for describing the behavior or decision making of multiple agents interacting with each other. It has been succesfully applied in numerous fields including biology \cite{molloy_inverse_2018}, economics \cite{dockner_differential_2000,carraro_tastes_1989} and automatic control. Within the control community, dynamic games have been studied and applied in the context of driver assistance systems \cite{flad_cooperative_2017}, multi-agent collision avoidance \cite{mylvaganam_differential_2017} and power system control \cite{chen_cooperative_2015}. In particular, several techniques for finding the optimal controls or decisions of each player based on known objectives have been thouroughly analyzed and applied. 

Recent years have seen a growing interest in the inverse problem of dynamic games, where the objectives modeled by cost or utility functions of each player are sought. This problem emerges when it is not possible to model objectives directly and it is desired to identify them based on previously observed actions of interacting players. These actions are typically assumed to correspond to a game equilibrium \cite{molloy_inverse_2018,carraro_tastes_1989,tsai_inverse_2016,rothfus_inverse_2017,molloy_inverse_2017-1,molloy_inverse_2020,inga_solution_2019,lin_multiagent_2018,reddy_inverse_2012,kopf_inverse_2017}. In this way, a general robust and transferable model (cf. \cite{ng_algorithms_2000}) of one or several agents in a multi-agent scenario can be obtained, e.g. human behavior in haptic dyad interaction \cite{inga_validation_2019} or bird collision avoidance behavior towards the design of unmanned aerial vehicle controllers \cite{molloy_inverse_2018}. This extends the \textit{learning by demonstration} paradigm to the multi-player case \cite{le_coordinated_2017}. Inspired by similar approaches in the single-agent scenario, also known as inverse optimal control, inverse dynamic game methods based on conditions for Nash equilibria have been proposed, e.g. for zero-sum games \cite{tsai_inverse_2016} or non-zero-sum two-player scenarios \cite{rothfus_inverse_2017}. Recent results in \cite{molloy_inverse_2017-1,molloy_inverse_2020, inga_solution_2019} show first extensions to a general N-player case.

The single-player problem has also been examined in the field of computer science, where various so-called inverse reinforcement learning (IRL) methods have been proposed (e.g. \cite{abbeel_apprenticeship_2004,zhou_infinite_2018}). Their focus is the identification of a cost function which may not be equal to the original one, but is able to explain observed trajectories. In the last years, some effort has been made to extend these techniques to a multiplayer setting. Many of these extensions consider a scenario where one global cost function is sought (e.g. of a central controller) which can describe the behavior of all agents \cite{hadfield-menell_cooperative_2016,natarajan_multi-agent_2010} or a scenario where all agents have the same reward \cite{sosic_inverse_2017}. These approaches are therefore related to cooperative dynamic games, where the players have an individual objective function but can cooperate in order to improve their performance\footnote{Both these state-of-the-art approaches and this paper consider cooperative yet not coalitional games, where several groups of players may build coalitions to act non-cooperatively with respect to other ones \cite{engwerda_positioning_2011}.}. As for non-cooperative dynamic games, some IRL-based methods have been proposed, e.g. \cite{lin_multiagent_2018, reddy_inverse_2012}. Nevertheless, all aforementioned IRL methods are based on a Markov Decision Process (MDP) and are limited to discrete-valued and finite control and state spaces. First endeavours of extending IRL methods to continuous-valued control and state spaces build upon the work in \cite{ziebart_maximum_2008} where the principle of maximum entropy (MaxEnt) \cite{jaynes_information_1957} was applied in a single-player IRL setting.  

Existing work in a multiplayer case includes \cite{kretzschmar_socially_2016} and \cite{ma_forecasting_2017}, where MaxEnt distributions are also considered. These papers as well as previous work in the single-player case \cite{kuderer_learning_2015,inga_individual_2017,inga_evaluating_2018} show the potential of MaxEnt IRL for real applications. However, the theoretical foundation of MaxEnt IRL in a multiplayer scenario has not been developed yet, especially in the case of continuous-valued state and control spaces. The latter is crucial to avoid the curse of dimensionality which would arise in many applications if discrete state and action spaces were assumed.

In this paper, we extend MaxEnt IRL to N-player inverse dynamic games with continuous-valued and infinite state and control spaces. We provide methods for identifying cost function parameters of one or several players in a dynamic game for three different solution concepts: (i) Pareto efficient solutions in cooperative games and (ii) open-loop and (iii) feedback Nash equilibrium solutions in non-cooperative games. Our approach for general N-player inverse non-cooperative dynamic games extends existing results in a single \cite{levine_continuous_2012} and two-player case \cite{kopf_inverse_2017}, where only linear-quadratic dynamic games with feedback Nash equilibrium solutions were considered. A further contribution of this paper are theoretical results which prove the unbiasedness of the estimation of the cost function parameters for each presented inverse dynamic game method.

Our paper is organized as follows. In Section \ref{sec:problem_def}, we define the general problem of inverse dynamic games. Then, the application of the principle of maximum entropy to an N-player dynamic game is shown in Section \ref{sec:MaxEnt_IDG}. A method for cost function parameter identification in cooperative games with Pareto efficient solutions is given in Section \ref{sec:Pareto}. Aftwerwards, in Section \ref{sec:NashGames} we propose an approach for non-cooperative inverse dynamic games with the open-loop and feedback Nash equilibrium solution concepts. The methods are illustrated with simulations in Section \ref{sec:SimulationExamples} before presenting conclusions in Section \ref{sec:Conclusion}.

\section{Problem Definition}\label{sec:problem_def}

Consider a dynamic game with $N$ players simultaneously controlling a (potentially time-variant) discrete-time system with the dynamics
\begin{equation}\label{system}
\m{x}^{(k+1)} = \m{f}^{(k)}(\m{x}^{(k)},\m{u}_1^{(k)},\m{u}_2^{(k)},\dots,\m{u}_N^{(k)})
\end{equation}
with the state values $\m{x}^{(k)} \in \mathbb{R}^{n}$ and the control values ${\m{u}_i^{(k)} \in \mathbb{R}^{m_i}}$, for all players $i \in \{1,\dots,N\}=:\mathbb{P} \subset \mathbb{N}^{+}$ and for all time steps ${k \in \{1,\dots,k_E\}=:\mathbb{K}\subset \mathbb{N}^{+}}$. The initial state ${\m{x}^{(1)}=\m{x}_1}$  is assumed to be known. The function ${\m{f}^{(k)}}$ is continuously differentiable with respect to all of its arguments ${\forall \, k \in \mathbb{K}}$. 
Each player $i \in \mathbb{P}$ minimizes his individual cost function $J_i$ by applying a sequence of control values $\m{u}_i^{(k)}, \forall k \in \mathbb{K}$. In this paper, we consider a widely used structure of the cost function which consists of a linear combination of $p_i \in \mathbb{N}^+$ known features (cf. \cite{molloy_inverse_2017-1} and references therein), i.e.
\begin{equation}\label{eq:cost_function}
J_i = -\sum_{k=1}^{k_E} \m{\theta}_i^{\top} \m{\eta}_i \left(\m{x}^{(k)}, \m{u}_1^{(k)}. \dots, \m{u}_N^{(k)} \right),
\end{equation}
where $\m{\eta}_i$ contains all $p_i$ features of player $i$ and $\m{\theta}_i \in \mathbb{R}^{p_i}$ represents the vector of player $i$'s individual parameters.
The features $\left(\eta_{i}\right)_{q}(\m{x}^{(k)},\m{u}_1^{(k)},\dots,\m{u}_N^{(k)})$, $q \in \{1,\dots,p_i\}$, of each player are assumed to be continuously differentiable with respect to all of their arguments for all $k \in \mathbb{K}$. 

Let $\underline{\m{x}} \in\mathbb{R}^{nk_E}$ and $\underline{\m{u}}_i \in\mathbb{R}^{m_i k_E}$, $i \in \mathbb{P}$, denote vectors containing all values of the system state~$\m{x}^{(k)}$ and the control values $\m{u}_i^{(k)}$ of player ${i \in \mathbb{P}}$ for all time steps $k \in \mathbb{K}$, respectively. With these, we define the following set:
\begin{definition}
A trajectory ${\zeta\coloneqq\left\{\underline{\m{x}}, \underline{\m{u}}_{1},  \dots, \underline{\m{u}}_{N}\right\}}$ is a set containing the values of the system state $\underline{\m{x}}$ and the controls $\underline{\m{u}}_i$ of all players ${i \in \mathbb{P}}$ which are feasible with respect to \eqref{system}.
\end{definition}
Observed trajectories are assumed to be generated by $\mathrm{p}\left(\zeta\right)$ which denotes a probability density function (PDF) over all possible trajectories $\zeta$. We assume that $n_t \in \mathbb{N}$ observations are available in the form of the observed trajectories~${\tilde{\zeta}\coloneqq\left\{\tilde{\underline{\m{x}}},\tilde{ \underline{\m{u}}}_{1},  \dots, \tilde{\underline{\m{u}}}_{N}\right\}}$ which belong to the set ${\mathcal{D} := \{\tilde{\zeta}_{1},\dots,\tilde{\zeta}_{{n_t}}\}}$. Each of these shall represent a solution of a dynamic game with cost functions $J_i$ parameterized by the unknown parameters $\m{\theta}_{i}^*$, ${\forall \, i \in \mathbb{P}}$. In the course of this paper, the solution will correspond to either Pareto efficient solutions, open-loop Nash or feedback Nash equilibrium solutions.

A key value in IRL methods is the feature count, which we introduce in the following.
\begin{definition}\label{def:feature_count}
The feature count $\m{\mu}_i\left(\zeta\right) \in \mathbb{R}^{p_i}$ of a player $i \in \mathbb{P}$ is defined as a vector containing the accumulated values of the features along a trajectory $\zeta$, i.e.
\begin{equation}\label{eq:feature_count}
\m{\mu}_i\left(\zeta\right) = \sum_{k=1}^{k_E} \m{\eta}_i\left(\m{x}^{(k)}, \m{u}_1^{(k)}, \dots, \m{u}_N^{(k)}\right),
\end{equation}
with ${\m{x}^{(k)}, \m{u}_i^{(k)} \in \zeta}$ , $\forall \, {i \in \mathbb{P}}$, ${k \in \mathbb{K}}$.  
\end{definition}
Using the feature counts $\m{\mu}_i\left(\zeta\right)$ and \eqref{eq:cost_function}, the costs along a trajectory $\zeta$ for any player $i \in \mathbb{P}$ can be rewritten as
\begin{equation}\label{eq:costs_feature_count}
	J_i\left(\zeta, \m{\theta}_i\right) = -\m{\theta}_i^{\top} \m{\mu}_i\left(\zeta\right).
\end{equation}
In the following, $\mathrm{p}\left(\left.\zeta\right|\m{\theta}_{1:N}\right)$ represents the probability density of a trajectory $\zeta$ which depends on the parameters $\m{\theta}_{1}, \dots, \m{\theta}_{N}$ of each player $i \in \mathbb{P}$.

An inverse dynamic game with IRL is defined as follows.
\begin{problem}\label{problem2}
	Find parameters $\hat{\m{\theta}}_i$, ${\forall \, i \in \mathbb{P}}$, such that the expected costs of a trajectory sampled from the resulting PDF $\mathrm{p}(\left.\zeta\right|\hat{\m{\theta}}_{1:N})$ corresponds for each player to the expected costs of a trajectory sampled from the PDF $\mathrm{p}\left(\left.\zeta\right|\m{\theta}_{1:N}^*\right)$, i.e. 
	\begin{equation}\label{eq:Forderung_theta_Kostenuebereinstimmung}
	\begin{split}
	&\mathbb{E}_{\mathrm{p}\left(\left.\zeta\right|\hat{\m{\theta}}_{1:N}\right)}\left\{J_{i}\left(\zeta, \m{\theta}_{i}^* \right)\right\} \overset{\text{!}}{=}\mathbb{E}_{\mathrm{p}\left(\left.\zeta\right|\m{\theta}_{1:N}^*\right)}\left\{J_{i}\left(\zeta, \m{\theta}_{i}^* \right)\right\}, \\
	\end{split}
	\end{equation}
		for all $i \in \mathbb{P}$.
\end{problem}	
The requirement \eqref{eq:Forderung_theta_Kostenuebereinstimmung} arises from the demand of obtaining for each player a cost function that results in an individual performance as good as the observed one, where the performance is measured with respect to each player's unknown true cost function. Without further assumptions, Problem \ref{problem2} is inherently ill-posed. This ill-posedness may be resolved by applying the principle of maximum entropy. In addition, including the knowledge of which solution concept lies at hand is necessary for the solution of an inverse dynamic game. These aspects shall be discussed in the next sections.


\section{Maximum Entropy For Inverse Dynamic Games}\label{sec:MaxEnt_IDG}

In this section, we apply the principle of maximum entropy to obtain a PDF which shall serve as a basis for identifying cost function parameters in $N$-player inverse dynamic games. This principle leads to the “least biased estimate possible on the given information”\cite{jaynes_information_1957}, where the information lies in the form of known moment constraints.\footnote{This is illustrated e.g. by the fact that the distribution which maximizes the entropy with the constraints of fixed and known expectation and variance is the Gaussian distribution. Furthermore, the maximum entropy distribution where no constraints are included is the uniform distribution \cite[Section 12.2]{cover_elements_2006}.} In order to to state a relationship between observed trajectories $\tilde{\zeta}$ and the expectation (first moment) of the observed trajectory feature count generated by the PDF $\mathrm{p}\left(\left.\zeta\right|\m{\theta}_{1:N}^*\right)$ which generated them, we make the following assumption.
\begin{assumption}\label{ass:featurecount_expectation}
The feature count along the observed trajectories in $\mathcal{D}$ represents the
expectation of the feature count $\mathbb{E}_{\mathrm{p}\left(\left.\zeta\right|\m{\theta}_{1:N}^*\right)}\left\{\m{\mu}_{i}\left(\zeta\right)\right\}$ based on the PDF $\mathrm{p}\left(\left.\zeta\right|\m{\theta}_{1:N}^*\right)$ which results from the original cost function parameters, i.e.
\begin{equation}\label{eq:global_merkmalswertExterte}
	\mathbb{E}_{\mathrm{p}\left(\left.\zeta\right|\m{\theta}_{1:N}^*\right)}\left\{\m{\mu}_{i}\left(\zeta\right)\right\} = \frac{1}{n_t} \sum_{l=1}^{n_t}\m{\mu}_{i}(\tilde{\zeta}_{l}) =: \tilde{\m{\mu}}_{i},
\end{equation}
where ${\m{\mu}}_i(\tilde{\zeta}_{l})$ denotes the feature count of the observed trajectory $\tilde{\zeta}_{l}$, $l \in \{1,...,n_t\}$.
\end{assumption}
Assumption \ref{ass:featurecount_expectation} implies that the observations are representative of the population consisting of all possible trajectories which can be generated from the PDF $\mathrm{p}(\left.\zeta\right|\m{\theta}_{1:N}^*)$. As no further information is available, the sample mean is used as an estimate for the expectation of the feature count.
\begin{lemma}\label{lem:featureweights_uebereinstimmung}
Let the expectation of the feature count be equal for both the PDFs $\mathrm{p}(\left.\zeta\right|\hat{\m{\theta}}_{1:N})$ and $\mathrm{p}\left(\left.\zeta\right|{\m{\theta}}_{1:N}^*\right)$, i.e.
\begin{equation}\label{eq:Forderung_theta_Merkmalsuebereinstimmung}
\mathbb{E}_{\mathrm{p}\left(\left.\zeta\right|\hat{\m{\theta}}_{1:N}\right)}\left\{ \m{\mu}_{i}\left(\zeta\right) \right\} = \mathbb{E}_{\mathrm{p}\left(\left.\zeta\right|\m{\theta}_{1:N}^*\right)}\left\{ \m{\mu}_{i}\left(\zeta\right) \right\},
\end{equation}
for each player $i \in \mathbb{P}$. Then, for any parameters $\m{\theta}_{i}^*$ with $\left\|\m{\theta}_{i}^* \right\|_{2} < \infty$, \eqref{eq:Forderung_theta_Kostenuebereinstimmung} is fulfilled.
\end{lemma}
\begin{IEEEproof}
The proof is analogous to the one-player case \cite{abbeel_apprenticeship_2004}.
\end{IEEEproof}

Lemma \ref{lem:featureweights_uebereinstimmung} represents the principle of matching feature expectations for all players. This principle was introduced in \cite{abbeel_apprenticeship_2004} and used as a basis for numerous single-player IRL methods.

Since Problem 1 implies \eqref{eq:Forderung_theta_Kostenuebereinstimmung}, by the results of Lemma \ref{lem:featureweights_uebereinstimmung} and using Assumption \ref{ass:featurecount_expectation} we 
require 
\begin{equation}\label{eq:matching_feature_counts}
\mathbb{E}_{\mathrm{p}\left(\left.\zeta\right|\hat{\m{\theta}}_{1:N}\right)}\left\{\m{\mu}_{i}\left(\zeta\right)\right\} = \tilde{\m{\mu}}_{i}, \quad\forall i \in \mathbb{P}.
\end{equation}
Our aim is to find a PDF $\mathrm{p}\left(\left.\zeta\right|\m{\theta}_{1:N}\right)$ which represents the probability of trajectories~$\zeta$ as a function of the parameters $\m{\theta}_1, \dots, \m{\theta}_N$, yet considering \eqref{eq:matching_feature_counts} as only a-priori knowledge.
Since this condition does not lead to a unique solution for the PDF, the principle of maximum entropy is applied. The following lemma states the PDF which maximizes the entropy within the framework of inverse dynamic games.
\begin{lemma}\label{lem:maxent_Nplayer}
	The unique maximum entropy PDF under the constraints defined by \eqref{eq:matching_feature_counts} is given by 
	\begin{equation}\label{eq:global_Wahrscheinlichkeitsdichte}
	\mathrm{p}\left(\left.\zeta\right|\m{\theta}_{1:N}\right) = \frac{\mathrm{exp}\left(\sum_{i=1}^{N} \m{\theta}_i^{\top}\m{\mu}_{i}\left(\zeta\right) \right)}{\int_{\forall{\zeta}} \mathrm{exp}\left(\sum_{i=1}^{N} \m{\theta}_i^{\top}\m{\mu}_i\left({\zeta}\right) \right) \mathrm{d}{\zeta}}.  \\
	\end{equation}
\end{lemma}

\begin{IEEEproof}
The lemma can be proved using a calculus-based approach analogously to \cite[Section 12.1]{cover_elements_2006}.
\end{IEEEproof}

%
%


In order to solve an inverse dynamic game based on the derived PDF, the knowledge of the underlying solution concept becomes necessary. The following section tackles this problem for cooperative games with Pareto efficient solutions.

\section{Inverse Cooperative Dynamic Games}\label{sec:Pareto}

In this section, we present a method to identify cost function parameters out of trajectories describing a Pareto efficient solution of the dynamic game. Furthermore, we prove the unbiasedness of the estimation.


\subsection{Preliminaries}

We restrict ourselves to Pareto efficient solutions which can be described by a global cost function given by the sum of weighted player cost functions. One particular global cost function is given by the sum of uniformly weighted player cost functions defined as follows.
\begin{definition}\label{def:globalcosts}
The uniformly weighted sum of all player cost functions is given by 
\begin{equation}\label{eq:globaleGuetefunktion}
J_{\Sigma} = \sum_{i=1}^{N} J_i = \sum_{i=1}^{N} -\m{\theta}_i^{\top}\m{\mu}_i =: -\m{\theta}_{\Sigma}^{\top}\m{\mu}_{\Sigma}
\end{equation}
with
\begin{IEEEeqnarray}{rCl}\IEEEyesnumber \label{eq:sum_variables}\IEEEyessubnumber
	\m{\theta}_{\Sigma} = \begin{bmatrix}
	\m{\theta}_1^{\top} & \dots & \m{\theta}_N^{\top}
	\end{bmatrix}^{\top}, \label{eq:sum_theta} \\   
	\noalign{\noindent and\vspace{\jot}} 
	\m{\mu}_{\Sigma} = \begin{bmatrix}
	\m{\mu}_1^{\top} & \dots & \m{\mu}_N^{\top}
	\end{bmatrix}^{\top}. \IEEEyessubnumber \label{eq:sum_mu} %
\end{IEEEeqnarray}
\end{definition}

We further introduce the following assumption.

\begin{assumption}\label{ass:convexJandGamma}
The cost functions $J_i$ are convex for all $i\in \mathbb{P}$.
\end{assumption}

\begin{remark}
We note that
\begin{equation}
\underset{\gamma}{\mathrm{arg~min~}} J_{\Sigma}(\gamma) = \underset{\gamma}{\mathrm{arg~min~}} \sum\limits_{i=1}^N \frac{1}{N} J_i(\gamma),
\end{equation}
where $\gamma := \{\underline{\m{u}}_1,\dots,\underline{\m{u}}_N\}$, holds since multiplying any cost function $J_{\Sigma}$ with a constant factor $c \in \mathbb{R}^+$ (here $\nicefrac{1}{N}$) does not alter the solution of the optimization problem. Therefore, under Assumption \ref{ass:convexJandGamma}, the minimizer of $J_{\Sigma}$ describes a Pareto efficient solution of a cooperative game \cite[Theorem 6.4]{engwerda_lq_2005}.
\end{remark}

\subsection{Identification Method and Unbiasedness of the Estimation}

 The method is based on the maximum likelihood of the observed trajectories under the PDF \eqref{eq:global_Wahrscheinlichkeitsdichte}. Before introducing the method, we use \eqref{eq:globaleGuetefunktion} and \eqref{eq:sum_variables} to rewrite \eqref{eq:global_Wahrscheinlichkeitsdichte} as
\begin{equation}\label{eq:global_Wahrscheinlichkeitsdichte2}
\begin{split}
\mathrm{p}\left(\left.\zeta\right|\m{\theta}_{\Sigma}\right) &= \frac{\text{exp}\left( \m{\theta}_{\Sigma}^{\top}\m{\mu}_{\Sigma}\left(\zeta\right) \right)}{\int_{\forall{\zeta}} \text{exp}\left( \m{\theta}_{\Sigma}^{\top}\m{\mu}_{\Sigma}({\zeta}) \right) \mathrm{d}{\zeta}}.
\end{split}
\end{equation}
The unbiased identification of cost functions in an inverse cooperative dynamic game is presented in the following theorem.
\begin{theorem}\label{thm:ErwartungstreueGlobal}
Let $n_t$ observed trajectories in $\mathcal{D}=\{\tilde{\zeta}_{1},\dots,\tilde{\zeta}_{n_t}\}$ fulfilling Assumption \ref{ass:featurecount_expectation} be available. Then, the maximum likelihood estimator (MLE) with respect to $\mathcal{D}$, i.e.
\begin{align}\label{eq:Ident_global_mehrereTraj_loglike}
\hat{\boldsymbol{\theta}}_{\Sigma}  &= \underset{\boldsymbol{\theta}_{\Sigma}}{\mathrm{arg~max~}} \mathrm{ln}~ \mathcal{L}\left\{\left.\boldsymbol{\theta}_{\Sigma}\right|\mathcal{D} \right\} \nonumber\\
&:=\underset{\boldsymbol{\theta}_{\Sigma}}{\mathrm{arg~max~}} \mathrm{ln}\prod_{l=1}^{n_t} \mathrm{p}\,(\tilde{\zeta}_{l}\,|\,\m{\theta}_{\Sigma}) 
\end{align}
where $\mathrm{p}\,(  \tilde{\zeta}_{l}  \,|\,\boldsymbol{\theta}_{\Sigma} )$ is obtained by evaluating \eqref{eq:global_Wahrscheinlichkeitsdichte2} with $\tilde{\zeta}_{l}$, $l \in \{1,...,n_t\}$, leads to a PDF for which the trajectories yield in expectation the same accumulated costs for all players as the trajectories corresponding to the PDF with original parameters, i.e.
	\begin{equation}\label{eq:Erwartungstreue_global_J}
	\mathbb{E}_{\mathrm{p}\left(\left.\zeta\right|\m{\theta}_{\Sigma}^* \right)} \left\{ J_{\Sigma}\left(\zeta,\m{\theta}_{\Sigma}^* \right) \right\} = \mathbb{E}_{\mathrm{p}\left(\left.\zeta\right|\hat{\m{\theta}}_{\Sigma} \right)} \left\{ J_{\Sigma}\left(\zeta,\m{\theta}_{\Sigma}^* \right) \right\}.
	\end{equation}
\end{theorem}

\begin{IEEEproof}
From \eqref{eq:Ident_global_mehrereTraj_loglike} we have
\begin{equation}\label{eq:Erwt_global}
	\m{0}  \overset{\text{!}}{=} \left. \frac{\partial}{\partial \m{\theta}_{\Sigma}} \sum_{l=1}^{n_t} \text{ln}~\mathrm{p}\,(\tilde{\zeta}_{l}\,|\,\m{\theta}_{\Sigma}) \right|_{\hat{\m{\theta}}_{\Sigma}}.
	\end{equation}
	Using the PDF of Lemma~\ref{lem:maxent_Nplayer}, this can be rewritten as
	{\small
	\begin{align}\label{eq:Erwartungstreue_global}
	\m{0}  &\overset{\text{!}}{=}	 \left. \frac{\partial}{\partial \m{\theta}_{\Sigma}} \sum_{l=1}^{n_t} \text{ln} \left( \frac{\text{exp}\left( \m{\theta}_{\Sigma}^{\top}\m{\mu}_{\Sigma}\left(\tilde{\zeta}_{l}\right) \right)}{\displaystyle\int_{{\zeta}} \text{exp}\left( \m{\theta}_{\Sigma}^{\top}\m{\mu}_{\Sigma}({\zeta}) \right)\mathrm{d}{\zeta} } \right) \right|_{\hat{\m{\theta}}_{\Sigma}}  \\
	&=  \left.\sum_{l=1}^{n_t} \frac{\partial}{\partial \m{\theta}_{\Sigma}} \hspace{-0.05cm} \left( \hspace{-0.08cm} -\text{ln}\left( \displaystyle\int_{{\zeta}} \text{exp}\left( \m{\theta}_{\Sigma}^{\top}\m{\mu}_{\Sigma}({\zeta}) \right)\mathrm{d}{\zeta} \right)\hspace{-0.04cm} + \hspace{-0.04cm}
	 \m{\theta}_{\Sigma}^{\top}\m{\mu}_{\Sigma}\left(\tilde{\zeta}_{l}\right) \hspace{-0.08cm} \right) \right|_{\hat{\m{\theta}}_{\Sigma}}   \\
	&= \left. \sum_{l=1}^{n_t} \Bigg( 
	\frac{\int_{{\zeta}} -\text{exp}\left( \m{\theta}_{\Sigma}^{\top}\m{\mu}_{\Sigma}({\zeta}) \right)\m{\mu}_{\Sigma}({\zeta}) \mathrm{d}{\zeta} }{\displaystyle\int_{\overline{\zeta}} \text{exp}\left( \m{\theta}_{\Sigma}^{\top}\m{\mu}_{\Sigma}\left({\overline{\zeta}}\right) \right)\mathrm{d}\overline{\zeta}} + 
	 \m{\mu}_{\Sigma}\left(\tilde{\zeta}_{l}\right)
	 \Bigg)\right|_{\hat{\m{\theta}}_{\Sigma}}  \label{eq:lasteq_erwartungstreue_global} 	 
	 \end{align}}
	 Since the integrals in the numerator and the denominator in \eqref{eq:lasteq_erwartungstreue_global} are independent of each other, \eqref{eq:lasteq_erwartungstreue_global} can be rewritten as
	 	 {\small
	 \begin{align}
\m{0}	&\overset{\text{!}}{=} \left. \sum_{l=1}^{n_t} \Bigg(\int_{{\zeta}} 
	\frac{-\text{exp}\left( \m{\theta}_{\Sigma}^{\top}\m{\mu}_{\Sigma}({\zeta}) \right)\m{\mu}_{\Sigma}({\zeta}) }{\displaystyle\int_{\overline{\zeta}} \text{exp}\left( \m{\theta}_{\Sigma}^{\top}\m{\mu}_{\Sigma}\left({\overline{\zeta}}\right) \right)\mathrm{d}\overline{\zeta} } \mathrm{d}{\zeta} +
	 \m{\mu}_{\Sigma}\left(\tilde{\zeta}_{l}\right) \Bigg) \right|_{\hat{\m{\theta}}_{\Sigma}}.  
	 \end{align}
	 }
	 Using \eqref{eq:global_Wahrscheinlichkeitsdichte2}, we get
	 \begin{align}
	\m{0}	&\overset{\text{!}}{=}   \sum_{l=1}^{n_t} \left(-\int_{{\zeta}} \mathrm{p}\left(\left.{\zeta}\right|\m{\theta}_{\Sigma} \right)\m{\mu}_{\Sigma}({\zeta}) \, \mathrm{d}{\zeta} + 
	\m{\mu}_{\Sigma}\left(\tilde{\zeta}_{l}\right)
	\bigg)\right|_{\hat{\m{\theta}}_{\Sigma}} \nonumber \\
	&= \sum_{l=1}^{n_t} \left(  -\mathbb{E}_{\mathrm{p}\left(\left.\zeta\right|\hat{\m{\theta}}_{\Sigma} \right)} \left\{ \m{\mu}_{\Sigma}(\zeta) \right\} + \m{\mu}_{\Sigma}\left(\tilde{\zeta}_{l}\right) \right). \label{eq:Erwartungstreue_global_last}
	\end{align}	
	From \eqref{eq:Erwartungstreue_global_last} and by Assumption \ref{ass:featurecount_expectation} we obtain	
	\begin{alignat}{2}\label{eq:Erwartungstreue_global_2}
&& \mathbb{E}_{\mathrm{p}\left(\left.\zeta\right|\hat{\m{\theta}}_{\Sigma} \right)} \left\{ \m{\mu}_{\Sigma}(\zeta) \right\} &= \frac{1}{n_t} \sum_{l=1}^{n_t} \m{\mu}_{\Sigma}\left(\tilde{\zeta}_{l}\right)  \\
	&& &= \mathbb{E}_{\mathrm{p}\left(\left.\zeta\right|\m{\theta}_{\Sigma}^* \right)} \left\{ \m{\mu}_{\Sigma}(\zeta) \right\}. \nonumber 
	\end{alignat}
	 By the results of Lemma \ref{lem:featureweights_uebereinstimmung}, \eqref{eq:Erwartungstreue_global_2} leads to \eqref{eq:Erwartungstreue_global_J}.
\end{IEEEproof}

Theorem \ref{thm:ErwartungstreueGlobal} implies that the expectation of the global costs (under the original parameters) produced by trajectories generated by the PDFs with original and estimated parameters are equal. While this result is generally weaker than the one required in \eqref{eq:Forderung_theta_Kostenuebereinstimmung}, it is enough to describe observed trajectories completely in cooperative games.

\begin{remark}\label{rem:levine}
Solving \eqref{eq:Ident_global_mehrereTraj_loglike} demands the possibility of evaluating $\mathcal{L}\left\{\left.\boldsymbol{\theta}_{\Sigma} \right|\mathcal{D}\right\}$ and therefore the PDF \eqref{eq:global_Wahrscheinlichkeitsdichte2} at the trajectories $\tilde{\zeta}_{l}$. Eq. \eqref{eq:global_Wahrscheinlichkeitsdichte2} includes an integral over all trajectories $\tilde{\zeta}$ which are feasible with respect to the system dynamics. Calculating this integral is intractable given the continuous-valued control and action spaces. Therefore, approximations are usually sought. In this paper, we apply the approach introduced in \cite{levine_continuous_2012} which involves a quadratic approximation of the cost function evaluated at the observed trajectory values. With this approach, the unbiasedness results hold exactly for quadratic cost functions. On the other hand, the exactness of the unbiasedness results depends on the optimality of the observed trajectories. Thus, if the observed trajectories correspond to an exact solution of the dynamic game, we have exact unbiasedness since these are also optimal with respect to the quadratic approximation.
\end{remark}
 
\begin{remark}\label{rem:par_recovery_CG}
Assuming that the number of features $p_i$ is known for all $i \in \mathbb{P}$, an individual parameter set $\hat{\m{\theta}}_i$ can be determined for each $i\in\mathbb{P}$ by means of \eqref{eq:sum_theta} out of the MLE $\hat{\m{\theta}}_{\Sigma}$.
\end{remark}
\begin{remark}\label{rem:Pareto}
Even though these results were derived by regarding uniformly weighted player cost functions, the presented method can also be used for explaining trajectories which arised from the sum of cost functions which are not necessarily equally weighted (see \cite[Definition 6.1]{engwerda_lq_2005}). 
\end{remark}


\section{Identification in Non-Cooperative Nash Games}\label{sec:NashGames}

We now consider non-cooperative dynamic games, where all players act greedily and no agreements between the players exist, leading to the Nash equilibrium solution concept.

\begin{definition} \label{def:Nash}
An $N$-tuple of control sequences $(\underline{\m{u}}^*_1,...,\underline{\m{u}}^*_N)$ constitutes a Nash equilibrium if, and only if, the inequality
\begin{equation}\label{eq:Nash}
J_i \left( \underline{\m{u}}_i^*, \underline{\m{u}}_{\lnot i}^*, \m{\theta}_i^* \right) \leq\, J_i \left( \underline{\m{u}}_i, \underline{\m{u}}_{\lnot i}^*, \m{\theta}_i^* \right)\\
\end{equation}
is satisfied for all players $i \in \mathbb{P}$, where $\underline{\m{u}}_{\lnot i}$ denotes the control sequence of all players except player $i$ (cf. \cite[p. 266]{basar_dynamic_1999}).
\end{definition}

%
In the following, we shall consider inverse problems with open-loop (OL) and memory-less perfect state (MPS) information patterns leading to open-loop or feedback Nash equilibrium solutions, respectively.\footnote{For a feedback Nash equilibrium, a further restriction needs to be added to \eqref{eq:Nash} (see \cite[Definition 6.2]{basar_dynamic_1999}). We used this definition but omitted it here due to space restrictions.}

\subsection{Inverse Open-Loop Dynamic Games}\label{subsec:OL_Nash}


We consider inverse open-loop dynamic games, where player strategies depend only on the initial state, i.e.	$\m{u}_i^{(k)} = \m{\gamma}_i^{(k)}(\m{x}^{(1)})$. Similar to Section \ref{sec:Pareto}, we seek a suitable PDF $\mathrm{p}(\zeta)$ for the estimation of cost function parameters. Inspired by \cite[Theorem~6.1]{basar_dynamic_1999}, where it can be discerned that the other players' controls do not have any direct influence on player $i$'s actions, we define the PDF 
\begin{align} \label{eq:WahrscheinlichkeitTrajektorieSpieleriOpenloop}
	\mathrm{p}\left(\left. \zeta \right| \m{\theta}_i \right) &= \frac{\mathrm{exp}\left( \m{\theta}_i^{\top}\m{\mu}_i(\zeta) \right)}{\displaystyle\int_{{\zeta}} \mathrm{exp}\left( \m{\theta}_i^{\top}\m{\mu}_i({{\zeta}}) \right)\mathrm{d}{\zeta} }
\end{align}
which represents the probability of a particular trajectory from the point of view of player $i$. This simplifies the PDF $\mathrm{p}\left(\left.\zeta\right|\m{\theta}_{1:N}\right)$ in such a way that $N$ PDFs $\mathrm{p}\left(\left. \zeta \right| \m{\theta}_i \right)$ which depend each on each player's cost function parameters $\m{\theta}_i$, $i \in \mathbb{P}$, are considered instead of one single PDF which depends on all parameters.

We introduce the following assumption which adapts Assumption \ref{ass:featurecount_expectation} to PDFs depending only on the parameters $\m{\theta}_i$ of one player $i \in \mathbb{P}$ as defined in \eqref{eq:WahrscheinlichkeitTrajektorieSpieleriOpenloop}.
\begin{assumption}\label{ass:featurecount_expectation_OL_Nash}
The mean of the feature count of the $n_t$ observed trajectories gives the expectation of the trajectory feature count resulting from \eqref{eq:WahrscheinlichkeitTrajektorieSpieleriOpenloop} with $\m{\theta}_i^*$, $i \in \mathbb{P}$, i.e.
	\begin{equation}\label{eq:Erwartungswert_Experte_Nash}
		\mathbb{E}_{\mathrm{p}\left(\left.\zeta\right|\m{\theta}_i^* \right)} \left\{ \m{\mu}_j(\zeta) \right\} = \frac{1}{n_t} \sum_{l=1}^{n_t} \m{\mu}_j\left(\tilde{\zeta}_{l}\right), \quad \forall i,j \in \mathbb{P}.  
	\end{equation} 
\end{assumption}
Furthermore, we present an alternative definition of the cost functions.
\begin{definition}\label{def:extended_par}
Let $\bar{\m{\eta}}$ denote an \textit{extended feature vector} which includes all features $(\eta_i)_{q}$ $\forall \, i \in \mathbb{P}$, $\forall \, q \in \{1,\dots, p_i\}$ of all $N$ players such that $(\bar{\eta})_r \neq (\bar{\eta})_s$ for all $r,s \in \{1,\dots,\mathrm{dim}(\bar{\m{\eta}})\}$ and $r \neq s$. 
The \textit{extended feature count} $\bar{\m{\mu}}(\zeta)$ is defined analogously according to Definition \ref{def:feature_count}. Furthermore, let the \textit{extended parameter vector} $\bar{\m{\theta}}_i$ be defined such that
\begin{equation}\label{eq:Guetefkt_ZusgefMerkmale}
J_i(\zeta) = \m{\theta}_i^{\top}\m{\mu}_i(\zeta) = \bar{\m{\theta}}_i^{\top}\bar{\m{\mu}}(\zeta), \quad \forall \, i \in \mathbb{P}.
\end{equation}
\end{definition}


The following theorem gives the main result of this section.

\begin{theorem}\label{theo:ErwartungstreueOL}

Let a set of trajectories $\mathcal{D}=\{\tilde{\zeta}_{1},\dots,\tilde{\zeta}_{n_t}\}$ be given such that Assumption \ref{ass:featurecount_expectation_OL_Nash} is fulfilled. Then, the MLE with respect to $\mathcal{D}$, i.e.
\begin{equation}\label{eq:Ident_openloop_allg_mehrereTraj_loglike}
	\hat{\boldsymbol{\theta}}_i  = \underset{\boldsymbol{\theta}_i}{\mathrm{arg~max~}} \mathrm{ln}~ \mathcal{L}\left\{\left.\boldsymbol{\theta}_i\right|\mathcal{D} \right\}
	=\underset{\boldsymbol{\theta}_i}{\mathrm{arg~max~}} \sum_{l=1}^{n_t} \mathrm{ln}\left(\mathrm{p}\,(  \tilde{\zeta}_{l} \,|\, \boldsymbol{\theta}_i )\right),
\end{equation}
where $\mathrm{p}\,(  \tilde{\zeta}_{l} \,|\, \boldsymbol{\theta}_i )$ is obtained by evaluating  \eqref{eq:WahrscheinlichkeitTrajektorieSpieleriOpenloop} with $\tilde{\zeta}_l$, $l \in \{1,...,n_t\}$, leads to parameters $\hat{\m{\theta}}_i$ such that \eqref{eq:WahrscheinlichkeitTrajektorieSpieleriOpenloop} results in an expectation of the cost function values $J_j\left(\zeta,\m{\theta}_j^* \right)$, $\forall j \in \mathbb{P}$ which is equal to the one corresponding to the PDF $\mathrm{p}\left(\left.\zeta\right|\m{\theta}_i^* \right)$, i.e.
	\begin{equation}\label{eq:Erwartungstreue_IRLopenloopNash}
		 \mathbb{E}_{\mathrm{p}\left(\left.\zeta\right|\hat{\m{\theta}}_i \right)} \left\{ J_j\left(\zeta,\m{\theta}_j^* \right) \right\}=  \mathbb{E}_{\mathrm{p}\left(\left.\zeta\right|\m{\theta}_i^* \right)} \left\{ J_j\left(\zeta,\m{\theta}_j^* \right) \right\} ,
	\end{equation}
	holds for all  $i,j \in \mathbb{P}$.
\end{theorem}
\vspace{0.2cm}
\begin{IEEEproof}
Using Definition \ref{def:extended_par}, \eqref{eq:Erwartungstreue_IRLopenloopNash} can be rewritten as
\begin{equation}\label{eq:Erwartungstreue_IRLopenloopNash_angepasst}
\mathbb{E}_{\mathrm{p}\left(\left.\zeta\right|\hat{\bar{\m{\theta}}}_i \right)} \left\{ J_j\left(\zeta,\bar{\m{\theta}}_j^* \right) \right\} = 	\mathbb{E}_{\mathrm{p}\left(\left.\zeta\right|\bar{\m{\theta}}_i^* \right)} \left\{ J_j\left(\zeta,\bar{\m{\theta}}_j^* \right) \right\} 
\end{equation}
for all $i,j \in \mathbb{P}$. The maximization in \eqref{eq:Ident_openloop_allg_mehrereTraj_loglike} implies
\begin{equation}\label{eq:Erwartungstreue2}
\begin{split}
	\m{0}  \overset{\text{!}}{=}	& \left. \frac{\partial}{\partial \bar{\m{\theta}}_i} \sum_{l=1}^{n_t} \text{ln} \left( \frac{\text{exp}\left( \bar{\m{\theta}}_i^{\top}\bar{\m{\mu}}(\tilde{\zeta}_{l}) \right)}{\displaystyle\int_{{\zeta}} \text{exp}\left( \bar{\m{\theta}}_i^{\top}\bar{\m{\mu}}({\zeta}) \right)\mathrm{d}{\zeta} } \right) \right|_{\bar{\m{\theta}}_i = \hat{\bar{\m{\theta}}}_i},
	\end{split}
\end{equation}
where we also used \eqref{eq:Guetefkt_ZusgefMerkmale}. The rest of the proof is similar to the proof of Theorem \ref{thm:ErwartungstreueGlobal}.
\end{IEEEproof}
The results of Theorem \ref{theo:ErwartungstreueOL} guarantee that the costs of the estimated and the original unknown cost functions are the same for all players, thus ensuring the fulfillment of \eqref{eq:Forderung_theta_Kostenuebereinstimmung} and solving Problem \ref{problem2} for open-loop dynamic games.
\begin{remark}
The evaluation of $\mathcal{L}\left\{\left.\boldsymbol{\theta}_i \right|\mathcal{D}\right\}$ at the trajectories $\tilde{\zeta}_{l}$ is done analogously to the previous section (see Remark~\ref{rem:levine}). The same holds for the results of the next subsection.
\end{remark}

\subsection{Inverse Feedback Nash Dynamic Games}\label{subsec:CL_IDG}
In this section, we give solutions for inverse dynamic games with the feedback Nash equilibrium as a solution concept. Therefore, we consider the MPS information structure given by $\m{u}_i^{(k)} = \m{\gamma}_i(\m{x}^{(k)})$.\footnote{According to \cite[p. 278]{basar_dynamic_1999}, the feedback Nash equilibrium solution under the MPS information pattern solely depends on $\m{x}^{(k)}$ at the time step $k$. The dependency on $\m{x}^{(1)}$ is given only for $k=1$.}
For the next results, the following additional assumption is needed.
\begin{assumption}\label{ass:control_laws}
The players' Nash equilibrium feedback control laws $\m{\gamma}_i^*$ are known or can be estimated from the observed trajectories available in $\mathcal{D}$.
\end{assumption}

\begin{remark}
Assumption \ref{ass:control_laws} is, in general, potentially restrictive. Indeed, even the forward problem of computing feedback Nash equilibrium strategies is difficult to solve. However, for linear-quadratic (LQ) dynamic games with linear feedback strategies
\begin{equation}\label{eq:linear_state_feedback}
\m{u}_i^{*(k)} = \m{\gamma}_i^*(\m{x}^{(k)}) = \m{K}_i^* \m{x}^{(k)},
\end{equation}
with $\m{K}_i^* \in \mathbb{R}^{m_i \times n}$ \cite[Section 8.3]{engwerda_lq_2005}, the estimation of $\m{K}_i^*$ can easily be performed with a least-squares approach, see e.g. \cite{inga_solution_2019}. A similar approach can potentially be applied for control-affine nonlinear systems with quadratic cost functions since the structure of the control strategy is also known in this case (cf. \cite[Lemma 1]{jiang_data-driven_2018}).
\end{remark}

If Assumption \ref{ass:control_laws} holds, the control laws of the players  $j \in \mathbb{P}$, $j \neq i$ can be used to rewrite \eqref{system} as
{\begin{align}\label{eq:Systemdynamik_closedloop}
	\m{x}^{(k+1)}&= \m{f}^{(k)}\left(\m{x}^{(k)}, \m{\gamma}_1^{*}\left( \m{x}^{(k)} \right) ,\dots,\m{u}_i^{(k)}, \dots, \m{\gamma}_N^{*}\left( \m{x}^{(k)} \right) \right) \nonumber\\
	&=: \m{f}_i^{(k)}\left(\m{x}^{(k)},\m{u}_i^{(k)}\right).
\end{align}}
In this way, it is possible for player $i$ to represent the system dynamics as a function of the system state $\m{x}$ and his own control variable $\m{u}_i$. Analogously, the features $\m{\eta}_i$ can be rewritten as:
\begin{equation}\label{eq:Nash_CL_features}
	\begin{split}
		\m{\eta}_i &= \m{\eta}_i(\m{x}^{(k)}, \m{u}_1^{(k)}, \dots, \m{u}_N^{(k)}) \\
		&= \m{\eta}_i \left( \m{x}^{(k)},\m{\gamma}_1^*\left( \m{x}^{(k)} \right) ,\dots,\m{u}_i^{(k)}, \dots, \m{\gamma}_N^*\left( \m{x}^{(k)} \right) \right) \\
		&= \m{\eta}_i \left( \m{x}^{(k)}, \m{u}_i^{(k)} \right).
	\end{split}
\end{equation}
Based on the representation \eqref{eq:Systemdynamik_closedloop} of the system dynamics from player $i$'s perspective and the rewritten features \eqref{eq:Nash_CL_features}, we state the following theorem.

\begin{theorem}\label{theo:ErwartungstreueCL}
Let a set of trajectories $\mathcal{D}=\{\tilde{\zeta}_{1},\dots,\tilde{\zeta}_{n_t}\}$ be given such that Assumption \ref{ass:featurecount_expectation_OL_Nash} is fulfilled. Furthermore, let Assumption \ref{ass:control_laws} hold. Then, the MLE with respect to $\mathcal{D}$, i.e.
\begin{equation}\label{eq:Ident_closedloop_allg_mehrereTraj_loglike}
	\hat{\boldsymbol{\theta}}_i  = \underset{\boldsymbol{\theta}_i}{\mathrm{arg~max~}} \mathrm{ln}~ \mathcal{L}\left\{\left.\boldsymbol{\theta}_i\right|\mathcal{D} \right\}
	=\underset{\boldsymbol{\theta}_i}{\mathrm{arg~max~}} \sum_{l=1}^{n_t} \mathrm{ln}\left(\mathrm{p}\,( \tilde{\zeta}_l \,|\, \boldsymbol{\theta}_i)\right),
\end{equation}
where $\mathrm{p}\,( \tilde{\zeta}_l \,|\, \boldsymbol{\theta}_i)$ is obtained by evaluating  \eqref{eq:WahrscheinlichkeitTrajektorieSpieleriOpenloop} with $\tilde{\zeta}_{l}$, $l \in \{1,...,n_t\}$ and with respect to \eqref{eq:Systemdynamik_closedloop}, leads to parameters $\hat{\m{\theta}}_i$ such that 
	\begin{equation}\label{eq:Erwartungstreue_IRLclosedloopNash}
	E_{\mathrm{p}\left(\left.\zeta\right|\hat{\m{\theta}}_i \right)} \left\{ J_j\left(\zeta,\m{\theta}_j^* \right) \right\}=	E_{\mathrm{p}\left(\left.\zeta\right|\m{\theta}_i^* \right)} \left\{ J_j\left(\zeta,\m{\theta}_j^* \right) \right\} 
	\end{equation}
	holds for all  $i,j \in \mathbb{P}$ (cf. Theorem \ref{theo:ErwartungstreueOL}).
\end{theorem}
\begin{IEEEproof}
Each $J_i$ can be rewritten using \eqref{eq:Nash_CL_features}. Afterwards, the proof is the same as in Theorem \ref{theo:ErwartungstreueOL}.
\end{IEEEproof}


\section{Simulation Examples}\label{sec:SimulationExamples} 

In this section, we present simulations to illustrate the inverse dynamic game methods. The first example is a nonlinear dynamic game, while the second involves an LQ dynamic game such that it is possible to calculate Nash equilibria in order to verify the inverse feedback Nash dynamic game solutions. We end this section with a discussion.

\subsection{Nonlinear Dynamic Game}\label{subsec:results_nonlinear}

We demonstrate the performance and compare the results of the approach presented in Section \ref{sec:Pareto} for cooperative games with Pareto efficient solutions and the method presented in Section \ref{subsec:OL_Nash} for OL Nash equilibria. The considered system is the well-known ball-on-beam system which is typically used for testing nonlinear controllers (e.g. \cite{hauser_nonlinear_1992}), yet controlled by two players simultaneously in this case.

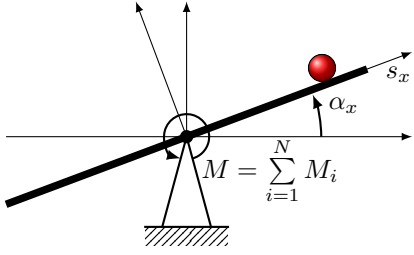
\begin{figure}[!t]
	\begin{center}	
		\usetikzlibrary{shapes,arrows}
\usetikzlibrary{calc} 
\usetikzlibrary{positioning}
\makeatletter
\pgfdeclareshape{datastore}{
  \inheritsavedanchors[from=rectangle]
  \inheritanchorborder[from=rectangle]
  \inheritanchor[from=rectangle]{center}
  \inheritanchor[from=rectangle]{base}
  \inheritanchor[from=rectangle]{north}
  \inheritanchor[from=rectangle]{north east}
  \inheritanchor[from=rectangle]{east}
  \inheritanchor[from=rectangle]{south east}
  \inheritanchor[from=rectangle]{south}
  \inheritanchor[from=rectangle]{south west}
  \inheritanchor[from=rectangle]{west}
  \inheritanchor[from=rectangle]{north west}
  \backgroundpath{
    \southwest \pgf@xa=\pgf@x \pgf@ya=\pgf@y
    \northeast \pgf@xb=\pgf@x \pgf@yb=\pgf@y
    \pgfpathmoveto{\pgfpoint{\pgf@xa}{\pgf@ya}}
    \pgfpathlineto{\pgfpoint{\pgf@xb}{\pgf@ya}}
    \pgfpathmoveto{\pgfpoint{\pgf@xa}{\pgf@yb}}
    \pgfpathlineto{\pgfpoint{\pgf@xb}{\pgf@yb}}
 }
}
\makeatother
\newcommand\DrawLongitudeCircle[2][1]{
	\LongitudePlane{\angEl}{#2}
	\tikzset{current plane/.prefix style={scale=#1}}
	\pgfmathsetmacro\angVis{atan(sin(#2)*cos(\angEl)/sin(\angEl))} %
	\draw[current plane] (\angVis:1) arc (\angVis:\angVis+180:1);
	\draw[current plane,dashed] (\angVis-180:1) arc (\angVis-180:\angVis:1);
}
\newcommand\DrawLatitudeCircle[2][1]{
	\LatitudePlane{\angEl}{#2}
	\tikzset{current plane/.prefix style={scale=#1}}
	\pgfmathsetmacro\sinVis{sin(#2)/cos(#2)*sin(\angEl)/cos(\angEl)}
	\pgfmathsetmacro\angVis{asin(min(1,max(\sinVis,-1)))}
	\draw[current plane] (\angVis:1) arc (\angVis:-\angVis-180:1);
	\draw[current plane,dashed] (180-\angVis:1) arc (180-\angVis:\angVis:1);
}
\tikzstyle{ground}=[fill,pattern=north east lines,draw=none,minimum width=1.1cm,minimum height=0.2cm]
	\begin{tikzpicture}[thick,>=latex',align=center, scale = 0.6]
	\node (ground1) [ground,yshift=-1.5cm,xshift=-1.25cm,anchor=north] {};
	\draw (ground1.north west) -- (ground1.north east);	
	\node (Boden_rechts) at($(ground1.north east)-(0.4,0)$){};
	\node (Boden_links) at($(ground1.north west)+(0.4,0)$){};
	\node (Lager) at($(ground1.north)+(0,2)$){};		
	\draw (Boden_rechts.center) -- (Lager.center);
	\draw (Boden_links.center) -- (Lager.center);
	\draw[fill=black](Lager.center)circle(3.5pt);
\draw[latex-] (Lager.center) ++(255:.5) arc (255:-75:.5)nodeat($(Lager.center)+(1.85cm, -0.75cm)$){$M = \sum\limits_{i=1}^N M_i$};
\node (Wippe_rechts) at($(Lager.center)+(4,1.5)$){};
\node (Wippe_links) at($(Lager.center)-(4,1.5)$){};
\draw[line width=0.1cm] (Wippe_links.center) -- (Wippe_rechts.center);
\node (Kugel) at($(Lager.center)+(3,1.524)$){};
\node (Achse_X) at($(Lager.center)+(5,0)$){};
\node (Achse_X_links) at($(Lager.center)+(-4,0)$){};
\draw[-latex,thin] (Achse_X_links.center) --  (Achse_X.center)node[anchor=north east]{};
\node (Achse_Z) at($(Lager.center)+(0,3)$){};
\draw[-latex,thin] (Lager.center) --  (Achse_Z.center)node[anchor=north east]{};
\node (achse_x) at($(Lager.center)+(5,1.875)$){};
\draw[-latex,thin] (Lager.center) --  (achse_x.center)node[anchor=north east]{};
\node (achsenbeschriftung_x) at($(achse_x.center)+(-0.3,-0.5)$){$s_x$};
\node (achse_z) at($(Lager.center)+(-1.5*3/4,3)$){};
\draw[-latex,thin] (Lager.center) --  (achse_z.center)node[anchor=north east]{};
\node (achsenbeschriftung_z) at($(achse_z.center)+(-0.2,-0.4)$){};
\draw[-latex] (0.9,-0.5) arc (0.5:22.5:2.5cm)nodeat(1.4,0.2){$\alpha_x$};
\def\R{.3} 
\def\angEl{35} 
\filldraw[ultra thin,ball color=red] (Kugel) circle (\R);
\end{tikzpicture}
	\end{center}	
\caption[Ball-on-Beam System]{Ball on beam system}
\label{fig:BallonBeam}
\end{figure}

The ball-on-beam system is depicted in Fig.~\ref{fig:BallonBeam}. Here, $\alpha_x$ denotes the angle of the beam towards the horizontal. In addition, $s_x$ represents the ball position in a beam-fixed coordinate system. Both players interact with the system by applying a torque $u_i(t)= M_i(t)$, $i \in \{1, 2\}$, with respect to the beam's rotational axis. Let the system state be defined as $\m{x}(t) = \begin{bmatrix} s_x(t) & \dot{s}_x(t) & \alpha_x(t) & \dot{\alpha}_x(t) \end{bmatrix}^\top$. Then, the system dynamics are described by the nonlinear differential equation
\begin{align}\label{eq:BoB_nonlinear}
\m{\dot{x}} = \begin{bmatrix}[1.2]
x_2 \\ \frac{m_b r_b^2 (x_1 x_4^2 - g \sin(x_3))}{\Theta_b + m_b r_b^2} \\ x_4 \\ \frac{-2m_b x_1x_2x_4 - m_bg x_1 \cos(x_3) + u_1 + u_2}{m_bx_1^2 + \Theta_p}
\end{bmatrix},
\end{align}
where $g$ denotes gravity, $\Theta_p$ is the inertia of the beam and $r_b$, $m_b$ and $\Theta_b$ are the radius, mass and inertia of the ball, respectively. All parameter values are given in Table \ref{tab:ball_on_beam_par}. 
 
\begin{table}[t] 
	\centering
	\caption{Parameters of the ball-on-beam system used for simulation}
	\begin{tabular}{ccccc}
		\toprule
		$g$ & $m_b$ & $r_b$  & $\Theta_b$ & $\Theta_p$  \\
		\midrule
		$\SI{9,81}{m/s^2}$ & $\SI{0.02}{\kilogram}$ & $\SI{25}{\milli \meter}$  & $\SI{5} \cdot {10^{-6}}\, \mathrm{kg}\,\mathrm{m}^2$ & $\SI{0.667}\,\mathrm{kg}\,\mathrm{m}^2$ \\
		\bottomrule
	\end{tabular}%
	\label{tab:ball_on_beam_par}
\end{table}%
In the following, units are neglected as all quantities are given in SI units. Each player acts based on an individual cost function of the form \eqref{eq:cost_function}, where the feature vector is given by
\begin{align}\label{eq:BoB_features}
\m{\eta}_i = -\begin{bmatrix} x_1^2 & x_2^2 & x_3^2 & x_4^2 & u_i^2 \end{bmatrix}^\top, \quad \forall i \in \{1,2\}.
\end{align}
The players' behavior is modeled by the parameters $\m{\theta}_{1}^* = \begin{bmatrix} 20 & 1 & 1 & 1 & 2 \end{bmatrix}^\top$ and $\m{\theta}_{2}^* = \begin{bmatrix} 1 & 1 & 10 & 1 & 1 \end{bmatrix}^\top$.

\subsubsection{Cooperative Game Solution}

We first assume that the players act cooperatively. Therefore, we determine optimal trajectories by solving an optimal control problem with the global cost function resulting from \eqref{eq:globaleGuetefunktion}, leading to
\begin{equation}
J_{\Sigma} = -\m{\theta}_{\Sigma}^\top \m{\mu}_{\Sigma} = -\begin{bmatrix} 21 & 2 & 11 & 2 & 3\end{bmatrix} \m{\mu}_i.
\end{equation}
This was done by applying Pontryagin's minimum principle and solving the resulting two-point boundary value problem (TPBVP). Using the initial state ${\m{x}^{(1)}=\begin{bmatrix}
0.5&0&0&0
\end{bmatrix}^\top}$, we obtain the observed trajectories $\tilde{\zeta}_{\text{CG}}$ of the cooperative game (CG) solution.

\subsubsection{Non-Cooperative Open-Loop Game Solution}

Similar to the CG solution, we apply Pontryagin's minimum principle and then solve the resulting TPBVP to determine the OL Nash equilibrium trajectories. The OL Nash equilibrium exists and is unique since the conditions of \cite[Lemma 4.2]{bressan_noncooperative_2011} are fulfilled. Using the same initial state $\m{x}^{(1)}$ as before, we obtain the observed $\tilde{\zeta}_{\text{NOLN}}$ corresponding to the (nonlinear) open-loop Nash equilibrium (NOLN).

\subsubsection{Inverse Dynamic Game Solutions}

\begin{figure}[!t]	
		\input{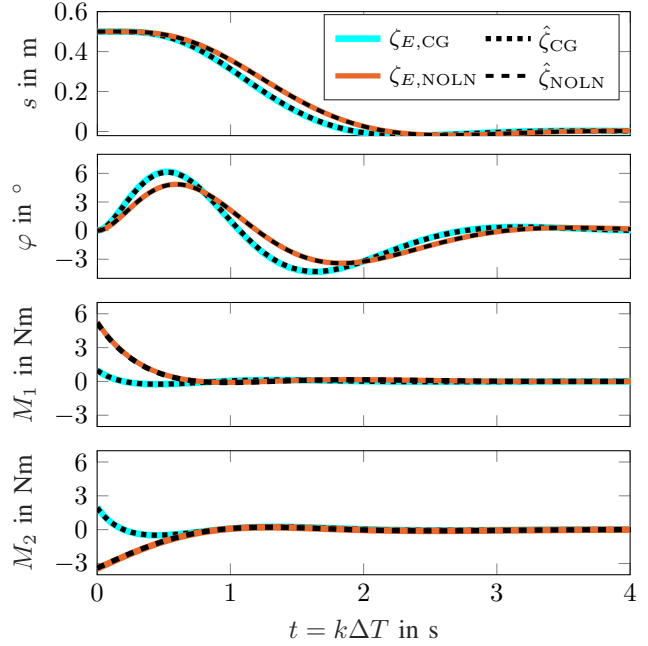}	
\caption{State and control trajectories used for identification (cooperative game and open-loop Nash solution) and corresponding trajectories generated from identified parameters}
\label{fig:nonlinear_results}
\end{figure}

In order to solve the inverse dynamic games corresponding to the Pareto and OL Nash solution concepts, the system was discretized using a sampling time $\Delta T = 0.02~\mathrm{s}$. The MLE \eqref{eq:Ident_global_mehrereTraj_loglike} was calculated with the Broyden-Fletcher-Goldfarb-Shanno (BFGS) method and the approach in Remark \ref{rem:par_recovery_CG} was applied to obtain the estimations $\m{\hat{\theta}}_i^{(\text{CG})}$. Similarly, \eqref{eq:Ident_openloop_allg_mehrereTraj_loglike} was solved to obtain the estimated parameters $\m{\hat{\theta}}_i^{(\text{NOLN})}$. All resulting parameters are given in Table \ref{tab:identification_par_NL}. These were used to generate estimated trajectories $\hat{\zeta}_{\text{CG}}$ and $\hat{\zeta}_{\text{NOLN}}$. All four sets of trajectories are depicted in Fig \ref{fig:nonlinear_results}. Error measures are given in Section \ref{subsec:results_noise}.

\begin{table}[t!]
\renewcommand{\arraystretch}{0.95}
\caption{Identified Parameters in the Nonlinear Dynamic Game}
\label{tab:identification_par_NL}
\centering
\begin{tabular}{cllllll}
\toprule

\multirow{3}{*}{$\m{\theta}_1$}& \textbf{*}    &  $[20.000$ & $1.000$ & $1.000$ & $1.000$ &$2.000]$ \\
								              &  \textbf{NOLN}  & 	$[19.697$ & $0.915$ & $2.350$ & $0.643$ & $2.000]$ \\
							                &	\textbf{CG}  & $[10.116$ & $1.207$ & $0.601$ & $1.317$ & $2.000]$ 	 \\

                               \midrule
\multirow{3}{*}{$\m{\theta}_2$}&\textbf{*}  &  $[1.000$ & $1.000$ & $10.000$ & $1.000$ & $1.000]$  \\
		   						 &  \textbf{NOLN}  &  $[1.027$ & $1.010$ & $\phantom{1}9.965$& $1.026$ & $1.000]$ \\
		   						 &  \textbf{CG}  &$[10.117$ & $1.203$ & $\phantom{1}0.601$ & $1.311$ & $1.000]$ \\
		   						                     
\bottomrule
\end{tabular}
\end{table}

\subsection{Linear-Quadratic Dynamic Game}

We now consider an LQ dynamic game to evaluate our approach in case of feedback (FB) Nash equilibria and compare it to the OL case. The cost functions $J_i$, ${i\in\left\{1, 2\right\}}$ have a form according to \eqref{eq:cost_function} and quadratic features given by \eqref{eq:BoB_features}. In order to obtain linear system dynamics, we linearize \eqref{eq:BoB_nonlinear} around $\m{x}(t) = \m{0}$, leading to
\begin{equation}\label{eq:BoB_linear}
\dot{\m{x}}(t) =  \m{A}\m{x}(t) + \m{B}_1 \m{u}_1(t) + \m{B}_2 \m{u}_2(t)
\end{equation}
where
\begin{equation}
\m{A}=\begin{bmatrix}
0&1&0&0\\0&0&\frac{-r_b^2m_bg}{\Theta_b+m_br_b^2}&0\\0&0&0&1\\\frac{-m_bg}{\Theta_p}&0&0&0
\end{bmatrix}, \, \m{B}_i=\begin{bmatrix}
0\\0\\0\\ \frac{1}{\Theta_p}
\end{bmatrix},
\end{equation}
$i\in\left\{1, 2\right\}$. Both parameter vectors $\m{\theta}_{1}^*$ and $\m{\theta}_{2}^*$ remain unchanged with respect to the previous example.

\subsubsection{Non-Cooperative Game Solution}

Using the ground truth cost function parameters, the feedback Nash equilibrium trajectories $\tilde{\zeta}_{\text{FB}}$ are calculated by means of the coupled matrix Riccati equations \cite[Theorem 8.5]{engwerda_lq_2005}. In order to give a comparison to another solution concept, we also calculate (linear) open-loop Nash (LOLN) equilibrium trajectories $\tilde{\zeta}_{\text{LOLN}}$ (based on \eqref{eq:BoB_linear}) analogously \cite[Theorem~7.13]{engwerda_lq_2005}. Both theorems allow to confirm the Nash character of the trajectories given the stability of the controlled system.

\subsubsection{Inverse Dynamic Game Solutions}

We discretize \eqref{eq:BoB_linear} using a sampling time $\Delta T = 0.02~\mathrm{s}$. Furthermore, for the FB Nash dynamic game, we previously estimated $\m{K}^*_i$ for both players using a least-squares approach based on \eqref{eq:linear_state_feedback} and the observed trajectories $\tilde{\zeta}_{\text{FB}}$.
Finally, the resulting optimization problem \eqref{eq:Ident_openloop_allg_mehrereTraj_loglike} is solved with the BFGS method for both the OL and FB Nash cases. The identified parameters are given in Table \ref{tab:identification_par_L}. These are used to determine the estimated trajectories $\hat{\zeta}_{\text{LOLN}}$ and $\hat{\zeta}_{\text{FB}}$. The observed and estimated trajectories are depicted in Fig. \ref{fig:LQ_results}. 

\subsection{Performance with Noisy Measurements}\label{subsec:results_noise}

\begin{table}[t!]
\renewcommand{\arraystretch}{0.95}
\caption{Identified Parameters in the LQ Dynamic Game}
\label{tab:identification_par_L}
\centering
\begin{tabular}{clc}
\toprule

\multirow{3}{*}{$\m{\theta}_1$} &  \textbf{*}  &  $\begin{bmatrix}20.000 & 1.000 & \phantom{-}1.000 & 1.000 &2.000 \end{bmatrix}$ \\
								&   \textbf{LOLN} &	$\begin{bmatrix}19.360 & 0.950 & \phantom{-}1.379 & 0.903 &2.000 \end{bmatrix}$ \\
								  &\textbf{FB} &	$\begin{bmatrix}19.421 & 1.002 & -0.375 & 1.017 &2.000 \end{bmatrix}$ \\
                               \midrule
\multirow{3}{*}{$\m{\theta}_2$} &  \textbf{*}  &  $\begin{bmatrix}1.000 & 1.000 & 10.000 & 1.000 & 1.000\end{bmatrix}$  \\
		   					&   \textbf{LOLN} &		 $\begin{bmatrix}0.640 & 0.928 & \phantom{1}9.531 & 0.962 & 1.000\end{bmatrix}$ \\
		   					&   \textbf{FB}   & $\begin{bmatrix}0.531 & 0.885 & \phantom{1}9.113 & 0.988 & 1.000\end{bmatrix}$  \\

\bottomrule
\end{tabular}
\end{table}

We now consider the case where measurements are imperfect. Gaussian noise is added to the states and controls to simulate noisy measurements $\tilde{x}_l(t) = \overline{x}_l(t) + \epsilon_l, \forall l \in \{1,...,n\}$ and $\tilde{u}_{i,k}(t) = \overline{u}_{i,k}(t) + \epsilon_{i,k}, \forall k \in \{1,...,m_i\}, \forall i\in\mathbb{P}$. Here, $\overline{\m{x}}(t)$ and $\overline{\m{u}}_i(t)$, $i \in \mathbb{P}$ stand for perfect observations of either Pareto efficient, OL or FB Nash equilibrium trajectories. The Gaussian noise is chosen such that all signals have a particular signal-to-noise ratio (SNR). We use different SNR levels for the evaluation. The performance of the methods is analyzed using the normalized maximum absolute error (NMAE) of the trajectories $e^{\m{a}} = \text{max }\{ e_j^{\m{a}}\}$ with
\begin{equation}
	e^{\boldsymbol{a}}_j = \left\| \frac{\hat{a}_j(t) - \overline{a}_j(t)}{\left\| \overline{a}_j(t) \right\|_{\text{max}}} \right\|_{\text{max}},
	\end{equation}
where $\m{a} \in \{\m{x}, \m{u}_i\}$ and $j \in \{1,\dots,n\}$ or $j \in \{1,...,m_i\}$ depending on whether state or controls are considered. We further consider $e^{\m{u}} = \text{max }\{ e^{\m{u}_i}\}$ for the controls. The results are given in Table \ref{tab:identification}. We denote with $\text{SNR} = \infty$ the case in which no noise is added to all signals.

\vspace{-0.05cm}

\subsection{Discussion}
\begin{figure}[!t]	
		\input{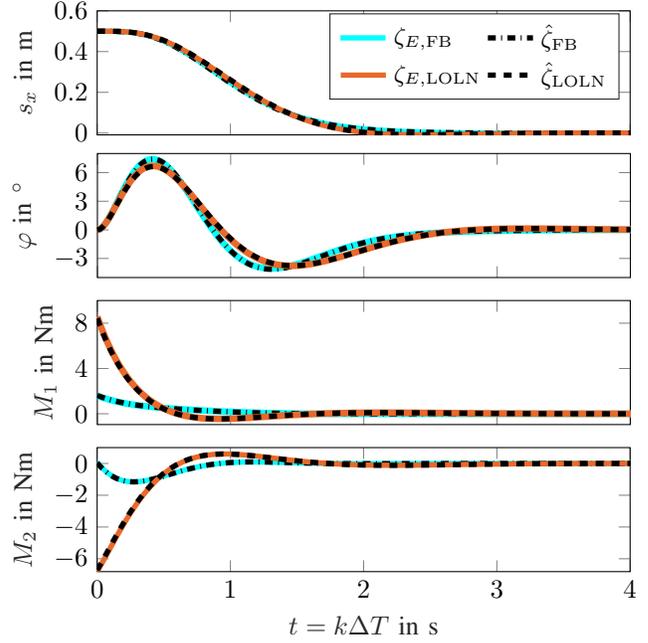}	
\caption{State and control trajectories used for identification (open-loop and feedback Nash solution) and corresponding trajectories generated from identified parameters}
\label{fig:LQ_results}
\end{figure}

We observe in Fig. \ref{fig:nonlinear_results} and Fig. \ref{fig:LQ_results} that the methods are able to determine cost function parameters which correctly explain the observed trajectories. For the CG case, we discern that the equally weighted sum associated to the identified individual parameters resembles the ground truth global cost function $J_{\Sigma}$. The correct approximation of the trajectories indicate that the identified parameters belong to the same Pareto frontier as the ground truth parameters. In addition, the individual parameters are almost equal for all common features (all except $u_i^2$) since the maximum entropy principle does not favour any player.

The identification of the CG solution described by the global cost function is robust to measurement noise. The other methods' results deteriorate for signals with an $\mathrm{SNR} < 20 \mathrm{dB}$. Furthermore, this effect grows with an increased number of maximum likelihood estimations in inverse Nash dynamic games.



\begin{table}[t!]
\renewcommand{\arraystretch}{0.9}
\caption{NMAE results with different SNR levels}
\label{tab:identification}
\centering
\begin{tabular}{ccccccc}
\toprule
\multicolumn{2}{c}{\textbf{SNR}}  & $15\, \mathrm{dB}$  &  $20\,\mathrm{dB}$  & $25\,\mathrm{dB}$  & $30\,\mathrm{dB}$ & $\infty$ \\
\midrule
\multirow{2}{*}{\textbf{CG}}  &  $e^{\m{x}}$  & 0.017 & 0.015 & 0.009 & 0.010 & 0.010 \\
                              &  $e^{\m{u}}$  & 0.016 & 0.017 & 0.005 & 0.006 & 0.003 \\
                              \midrule
\multirow{2}{*}{\textbf{NOLN}}&  $e^{\m{x}}$  & 0.041 & 0.019 & 0.009 & 0.005 & 0.003 \\
                              &  $e^{\m{u}}$  & 0.614 & 0.288 & 0.089 & 0.050 & 0.014 \\  
                              \midrule  
\multirow{2}{*}{\textbf{LOLN}}&  $e^{\m{x}}$  & 0.055 & 0.036 & 0.022 & 0.017 & 0.012 \\
                              &  $e^{\m{u}}$  & 1.046 & 0.375 & 0.319 & 0.016 & 0.004 \\                           
                              \midrule
\multirow{2}{*}{\textbf{FB}}  &  $e^{\m{x}}$  & 0.359 & 0.301 & 0.071 & 0.025 & 0.013 \\
                              &  $e^{\m{u}}$  & 0.998 & 0.382 & 0.144 & 0.032 & 0.032 \\
                              
\bottomrule
\end{tabular}
\end{table}

\section{Conclusion}\label{sec:Conclusion}

We developed methods for inverse dynamic games based on MaxEnt IRL, treating three different solution concepts and presenting unbiasedness results for case. The performance of the methods was shown using examples of nonlinear and LQ dynamic games.
Our methods allow for cost function identification in dynamic games to obtain a model of the players based on observed data. This can be done either by a centralized approach or by the agents themselves in case complete trajectory sets can be determined. For a more efficient application of these approaches, methods for the online computation of the likelihood function based on potentially incomplete trajectory sets are yet to be investigated in future work.


%





\ifCLASSOPTIONcaptionsoff
  \newpage
\fi




\bibliographystyle{IEEEtran}

\bibliography{Library}
\end{document}